# MAIN INJECTOR LCW (Low Conductivity Water) CONTROL SYSTEM

K. C. Seino, FNAL, Batavia, IL 60510, USA

Abstract

There are six service buildings uniformly spaced along the perimeter of MI (Main Injector). A total of 18 LCW pumps were installed around the MI ring with 3 pumps per building. Approximately 8,000 GPM of LCW is required to cool magnets, bus and power supplies in the MI enclosure and service buildings. In each service building, a PLC control system controls pumps and valves, and it monitors pressures, flow, resistivities and temperatures. The PLC hardware system consists of a Gateway module and a variety of I/O modules, which are made by Sixnet of Clifton Park, NY. The control system communicates with other buildings including MCR (Main Control Room) via an Ethernet link and front-end computers. For more details of the MI LCW control system, refer to [1] and [2]. One of the key elements of the PLC software is called ISaGRAF workbench, which was created by CJ International of Seyssins, France. The workbench provides a comprehensive control-programming environment, where control programs can be written in five different languages. For more details of ISaGRAF, refer to [3].

## 1 HARDWARE

### 1.1 PLC

The PLC hardware system consists of a Gateway module (local controller/remote communicator) and a variety of I/O modules. The product line of these modules is called as Sixtrak and is made by Sixnet of Clifton Park, NY. The Gateway module can be connected up to 20 I/O modules without expander modules, and it can be connected up to 128 I/O modules with expanders. The Gateway scans I/O modules and updates the information on them under a program control, and it communicates with host computers via an Ethernet link. The Gateway has 1 Mbytes of flash memory (for firmware), 256 Kb of battery backed RAM (for data), a real time clock and a RS232 serial port for local diagnostics.

For our applications, we are using several different types of I/O modules. They are analog input modules (4-20ma input, instrumentation and RTD), analog output modules, digital input modules (5VDC and 24VDC) and digital output modules (relay).

The modules are DIN rail mountable for instantaneous installations, and they can be removed from their wiring bases for easy installation and maintenance. Moreover, every module is isolated from a common bus and other modules for fault-free operations.

### 1.2 Instrumentation

The sensors, transmitters, local indicators, actuators and other field devices had been specified, purchased and installed by the Mechanical Support group. However, we did cabling/wiring, set key parameters, calibrated, did some adjustments and did trouble-shooting on these field devices. For this reason, I will briefly make reference to them as follows.

(1) Pressure Transducers (Measurement Specialties MSP-400-P-0064A, 0-400 psig), (2) Flow Transmitters (Peek Measurement 2120R, 0-100 GPM), (3) Differential Pressure Transmitters (Rosemount 2024), (4) Liquid Level Indicators (Gems SureSite 86158), (5) Resistivity Meters (Thornton 200CR, 0.0013-50 Mohm-cm with 0.1 constant cell), (6) RTD Probes (Devar #RTDSF2.5), (7) Electric Valve Actuators (Keystone EPI-13, 1/4 turn (close/open)), (8) Electric Valve Actuators for Temperature Regulation (EIM 2000 M2CP, 1/4 turn (continuous)), and (9) Controller for EIM2000 above (Powers 535, PID control).

### 1.3 Cables

The cables were pulled in MI<10:50>, 52 & 60 in 10/97 and 11/97. Since then, we had spent about ten months on an inconstant basis to terminate the cables, doing tests and fixing problems on the cables, the instrumentation and the PLC hardware for a total of nine systems excluding one for CUB.

We used several different types of cables -- (a) Belden #8761 for general use, (b) Belden #8719 for high voltage, (c) Belden #9533 for RTD, (d) Belden #9886 for 450 FT travel from MI-40 service building to Beam Dump, (e) Belden #8760 for Beam Dump, (f) Omega KK-J-20 for Type-J TCs (thermocouples) and (g) Omega KK-K-20 for Type-K TCs.

## 2 SOFTWARE

## 2.1 Sixnet Plant Floor

Plant Floor is a configuration and maintenance tool for Sixtrak I/O systems. Using Plant Floor's windows and menus, one assembles a graphic representation of each I/O system. Configuration choices let him customize each module in a given system. Once a system configuration is complete, the configuration is downloaded to the Gateway module of the system. Plant Floor is also a calibration tool for analog I/O modules, and it provides real-time displays for the maintenance and diagnostics on I/O modules.

## 2.2 ISaGRAF

ISaGRAF is a comprehensive control-programming environment that makes Sixtrak I/O a high performance, yet inexpensive controller. ISaGRAF uses standard industrial PLC programming methodologies for designing powerful applications without the need of high-level computer languages. ISaGRAF is created by CJ International of Seyssins, France, and it is sold by Sixnet as a part of the Sixnet software package.

An ISaGRAF project is a collection of individual programs and functions that form a complete control application. Each program controls one particular part of the application. In February 1993, responding to the need for standards to reduce training costs and guarantee portability, the IEC issued the IEC 1131-3 standard: a specification of five PLC languages that can be mixed in the same application. The five languages are (a) Sequential Function Chart (SFC), (b) Function Block Diagram (FBD), (c) Ladder Diagram (LD), (d) Structured Text (ST) and (e) Instruction List (IL).

An ISaGRAF project consists of programs, sub-programs and functions, which are placed in four different sections as follows.
(1) Beginning: Programs in this section are systematically executed in the beginning of a cycle after updating external inputs and outputs.
(2) Sequential: Programs in this section are executed confirming to the SFC rules and the implementation on a series of steps and transitions.
(3) End: Programs are executed at the end of a cycle just before updating external outputs.
(4) Functions: Sub-programs which can be called by programs in any of the other three sections.

The ISaGRAF dictionary is simply the collection of internal, input and output variables and defines that are used in the programs of a project. Variables and defines are specified as Local (specific to one program), Global (in any program within a project) or Common (in any project in ISaGRAF), when they are created in the dictionary. The Sixtags utility (a shared tag database): The I/O and module tag names that are created in the Plant Floor configuration can be exported to the dictionary of an ISaGRAF project by the Export command of Sixtags.

The ISaGRAF I/O simulator can be run by clicking on Simulate in the Debugger menu. The simulator allows us to try out a program before it is run on a live system. This convenient tool saves time by discovering problems and fixing them before a real start-up. I/O variables can be locked (disconnected) from their corresponding external devices. Once they are locked, their status/values can be altered by the debugger to proceed with the simulation.

Fig. 1 shows an example of the sequential program. For further details on ISaGRAF, one should refer to Ref. [1].

## 2.3 Interlocks/Trips

A number of analog and digital signals are examined on the LCW control systems, and if certain conditions exist, the programs use them to interlock the pump/valve operations. For each of the above analog signals, upper and lower trip limits are specified, and if the signal exceeds these limits, the program interlocks the related devices.

At two times (Start-up and Run times), conditions are tested. At the start-up time, if certain conditions exist, the LCW pumps can not be turned on by the console command, and the trips are set true. In this situation, the operator firstly removes the conditions and clears the trips. Secondly he sends the second command from the console to turn on the LCW pumps, because the first command has been cleared when the trips occurred.

During the run time, if certain conditions occur, the LCW pumps are turned off, and the trips are set true. In this situation, the operator follows the same procedure mentioned above.

There are two special rules to observe. One is 'Initial 15 sec Delay'. For 15 sec after the start-up, some conditions are ignored for tripping. The other is '15 sec Filtering'. Signals are sampled at a certain rate. If the conditions exist when sampled, the occurrences are counted. If the counts exceed their specified limits over the 15 sec periods, the LCW pumps are tripped off.

## 2.4 ACNET Connections

In each of the service buildings (MI-10, 20, 30, 40, 50, 52 & 60), the Gateway module communicates with the front-end computer of the house via an Ethernet link. The front-end in turn communicates with the console system via the Ethernet link.

There are a total of 9 front-ends that handle tasks related to the LCW controls. Some of them perform special tasks as well as communication. The MI-60 front-end performs a number of special tasks -- (a) it reads the status on four local valves and writes it at CUB for MI Makeup, (b) it reads the status on M_V03 (MI-60 Magnet V03) and writes it to all the other locations, and (c) it reads the status of the LCW pumps around the MI ring, examines it for a condition (if all the pumps are off or not) and uses the result for an action, and (d) it also writes the result to CUB. The front-ends at MI-52, MI-60 and CUB do calculations to convert the liquid levels to their volume equivalents.

## 2.5 Console Pages

The I56 page and its subpages show all the activities of the MI LCW system. The Global subpage shows an overall view of the pump status/controls, where the operator can view the status on LCW pumps, Pond pumps, transfer pumps and other pumps, where he can turn on pond pumps individually, and where he can turn on LCW pumps using a list (specifying the turn-on order and pauses between).

The MI-10 subpage shows a view of the MI-10 LCW system, which displays the status/controls on pumps, valves and PLC, analog readings, trip flags and limits. Fig. 2 is a graphic view of the MI-10 LCW system.

Fig. 1: Example of Sequential Program

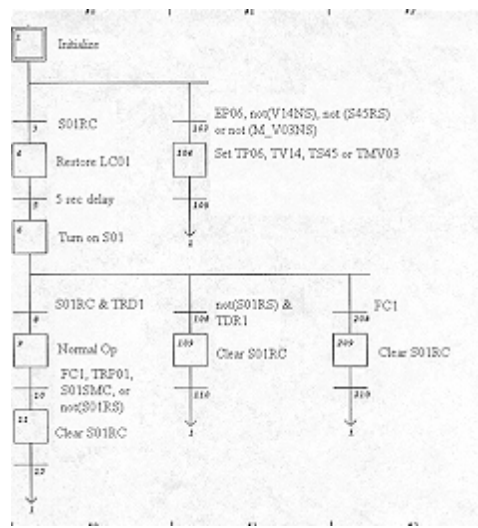

Fig. 2: Graphic View of MI-10 LCW System

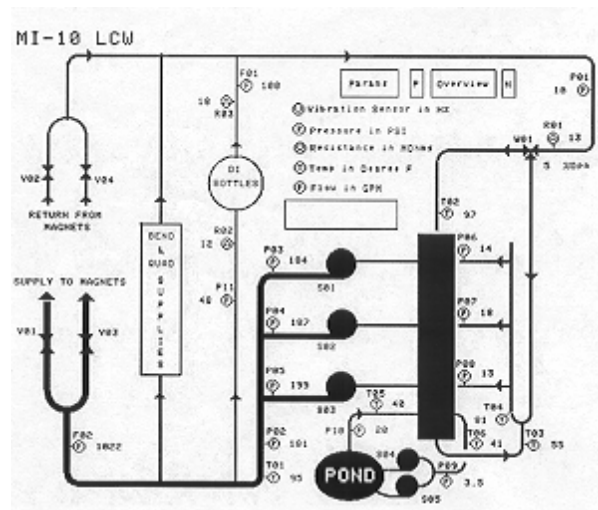